\documentclass[pra,aps,twocolumn,showpacs]{revtex4}
\usepackage{graphicx}
\usepackage{amsmath}

\begin{document}
\title{Fine tuning of quantum operations performed via Raman transitions}

\author{Grzegorz Chimczak and Ryszard Tana\'s}

\affiliation{Nonlinear Optics Division, Institute of Physics, Adam
  Mickiewicz University, 61-614 Pozna\'n, Poland}

\date{\today} \email{chimczak@kielich.amu.edu.pl}


\begin{abstract}
  A scheme for fine tuning of quantum operations to improve their
  performance is proposed.  A quantum system in $\Lambda$
  configuration with two-photon Raman transitions is considered
  without adiabatic elimination of the excited (intermediate) state.
  Conditional dynamics of the system is studied with focus on
  improving fidelity of quantum operations.  In particular, the $\pi$
  pulse and $\pi/2$ pulse quantum operations are considered.  The
  dressed states for the atom-field system, with an atom driven on one
  transition by a classical field and on the other by a quantum cavity
  field, are found. A discrete set of detunings is given for which
  high fidelity of desired states is achieved.  Analytical solutions for the
  quantum state amplitudes are found in the first order perturbation
  theory with respect to the cavity damping rate $\kappa$ and the
  spontaneous emission rate $\gamma$.  Numerical solutions for higher
  values of $\kappa$ and $\gamma$ indicate a stabilizing role of
  spontaneous emission in the $\pi$ and $\pi/2$ pulse quantum
  operations. The idea can also be applied for excitation pulses of
  different shapes. 
\end{abstract}

\maketitle
\section{Introduction}
An atomic $\Lambda$ system with Zeeman sublevels of the ground state,
which are sufficiently long lived to store qubits, plays a very
important role in quantum computations.  Superpositions of such states
can store quantum information for sufficiently long time to make
various quantum algorithms feasible~\cite{langer05_10s}.  Therefore,
researchers frequently consider such states in their
proposals~\cite{parkins93:_synth_zeeman,pell95:_decoh,bose,beige,
chimczak:_entanglement,lim06:_repeat,yuanJPB07}
and use them in their experiments~\cite{riebe04,barrett04,
mckeever04:_single_photon,boozerPRL07_map,legero04,hijlkemaNatPhy07_1fotSer}.
Access to a good quantum memory is necessary, but not sufficient, to
realize quantum computation. It is also important to be able to
perform quantum operations on such coded qubits with high fidelity
differing from unity by $10^{-5}$ or
less~\cite{preskill,steane99_doklad}.

Although there is no direct, strong coupling between the Zeeman
sublevels, one can efficiently manipulate populations of the states
using the two-photon Raman transition involving an auxiliary level,
for example, state $|2\rangle$ in the $\Lambda$ configuration, as
shown in Fig.~1. 
This method of qubit encoding and manipulation has
many different implementations. There are single atoms or ions modeled
by different level configuration systems: a three-level $\Lambda$
system~\cite{legero04,boozerPRL07_map}, a four-level system with three
levels in $\Lambda$ configuration, and one additional long-lived level,
which does not participate in the transition~\cite{Enk97,pachos02}, a
six-level double $\Lambda$ system, which consists of two $\Lambda$
systems behaving exactly in
parallel~\cite{pell95:_decoh,lim06:_repeat,schonPRA07}.  There are
also solid-state implementations: quantum dots modeled by the
$\Lambda$-type~\cite{kirazPRA04, djuricPRB07} three-level system and
superconducting quantum interference devices modeled by the 
$\Lambda$-type three-level
system~\cite{zhou02_SQUIDLam,amin03_SQUIDLam,yang03_SQUIDLam,
  yang04_SQUIDLam,yangPRL04_SQUIDLam}. In all these systems it is easy
to store qubits and it is easy to perform single qubit gates just by
turning the lasers on and off, to drive transitions for a proper
period of time. Moreover, $\Lambda$-type systems are also perfect to
realize the multiple qubit gates, as for example the crucial for
quantum computation controlled NOT gate~\cite{beige,tregenna02_cnot,gotoPRA04},
because one can place two or more such systems in a microcavity and
couple one transition in each $\Lambda$-type structure to the cavity
field mode.  Then, qubit interactions are mediated by the cavity field
mode.  Another way to realize two qubit gates is to perform the joint
detection of photons leaking out of two separate cavities with trapped
$\Lambda$-type systems~\cite{lim06:_repeat}.

Since there are many advantages of devices composed of $\Lambda$-type
systems and cavities, such systems are very popular elements of
various quantum information processing schemes.  However, there is one
important drawback of such schemes --- the population transferred to
the intermediate level diminishes the fidelity of quantum operations
performed with them. To avoid the destructive effect of populating the
intermediate level, one can decide to work in the decoherence-free
space~\cite{pell95:_decoh,gotoPRA04,beige,tregenna02_cnot,
  kis02:_qubit_rotat_by_stimul_raman_adiab_passag,
  sangouard05:_fast_swap_gate_by_adiab_passag} or, alternatively, one
can assume that the cavity mode and the laser fields are far detuned
from their respective transitions, as to make the population of the
intermediate state negligible. The latter approach is referred to as
adiabatic elimination.  In realistic situations with not extremely
large values of detunings, the population of the intermediate level is
small but noticeable enough to significantly deteriorate the quality of
quantum operations.
\begin{figure}[htbp]
  \centering
  \includegraphics[width=5cm]{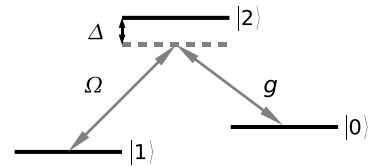}
  \caption{$\Lambda$-type Raman transition: a key ingredient of
    different quantum devices.}
  \label{fig:LV}
\end{figure}

In this paper, we propose a scheme for fine tuning of the quantum
operations performed via Raman transitions in the atomic $\Lambda$ system
driven on one transition by a classical field, and on the other by a
quantum cavity field. We discuss the conditional evolution of the
$\Lambda$ system in a situation when the adiabatic elimination is not
made, and we show that it is possible to take advantage of the fact
that the intermediate state population oscillates rapidly,
periodically approaching zero, which can be used to improve the
quality of quantum gates based on the Raman transitions in the
$\Lambda$ system. It turns out that it is possible, by setting
appropriate values for the detuning, to make the population of the
intermediate state negligible at the time the quantum operation is
completed, and in this way we are able to increase significantly the
fidelity of the operation. Such fine tuning of the quantum operations works
well even for small detunings and, therefore, it can prove useful when
one wants to perform quantum operations with high fidelity. We have
found a discrete set of detunings for which perfect operations are
possible if there is no damping. We have also found approximate
analytical solutions describing the evolution of the $\Lambda$ system
including both the cavity damping and spontaneous emission. Numerical
results for higher values of the cavity decay rate and spontaneous
emission rate show, somewhat unexpectedly, that the spontaneous
emission can play a stabilizing role improving the result of quantum
operation when the cavity decay rate becomes sufficiently large. We shortly address
the issue of fine tuning for the nonrectangular pulses.

\section{Model}
Let us consider two basic operations, which can be performed on
qubit-encoding states of the $\Lambda$-type level structures: the
$\pi$ pulse operation and the $\pi/2$ pulse operation. These two basic
operations can be used to realize a number of important quantum
information tasks as, for example, generation of maximally entangled
states~\cite{chimczak:_entanglement}, quantum information
transfer~\cite{bose,chimczak:_entanglement_teleportation}, or
the performing of controlled two qubit gates~\cite{lim06:_repeat}.
We consider an atom in $\Lambda$ configuration trapped in a cavity.
The long-lived states $|0\rangle$ and $|1\rangle$ of the atom are
coupled via the intermediate level $|2\rangle$ (see
Fig.~\ref{fig:LV}).  The $|0\rangle\leftrightarrow|2\rangle$
transition is coupled to the cavity mode with a frequency
$\omega_{\textrm{cav}}$ and coupling strength $g$. The second
transition is driven by a classical laser field with the coupling
strength $\Omega$.  The frequency of the laser field is
$\omega_{\textrm{L}}$.  Both the classical laser field and the
quantized cavity mode are detuned from the corresponding transition
frequencies by $\Delta=(E_{2}-E_{1})/\hbar -\omega_{L}$.  The
population transfer between long-lived states $|0\rangle$ and
$|1\rangle$ take place only when both transitions
$|0\rangle\leftrightarrow|2\rangle$ and
$|1\rangle\leftrightarrow|2\rangle$ are driven.  Therefore, the qubit
is safe if the laser is turned off. If we want to perform the $\pi$
pulse operation or the $\pi/2$ pulse operation then all we need to do
is turning the laser on for a proper period of time.

The evolution of the $\Lambda$-type quantum systems is determined by
the effective non-Hermitian Hamiltonian (we set $\hbar=1$ here and in
the following):
\begin{eqnarray}
  \label{eq:Hamiltonian0L}
  H =(\Delta-i\gamma) \sigma_{22} 
  +(\Omega \sigma_{21}+g a \sigma_{20}+ {\rm{H.c.}})
  -i \kappa a^{\dagger} a \, ,
\end{eqnarray}
where $\gamma$ is the spontaneous emission rate from the atomic state
$|2\rangle$, and $\kappa$ is the cavity decay rate.  In
expression~(\ref{eq:Hamiltonian0L}) we also introduce the flip
operators $\sigma_{ij}=|i\rangle\langle j|$, where \mbox{$i,j=0,1,2$}.
The effective Hamiltonian~(\ref{eq:Hamiltonian0L}) describes
conditional evolution of the atom-field system and will be used in our
further calculations.

\section{Exact solutions}
As a first step, we find the solutions for the Schr\"odinger equation
when both $\gamma$ and $\kappa$ are zeros. This assumption allows us
to obtain exact solutions. The Hamiltonian~(\ref{eq:Hamiltonian0L})
takes in this case the form
\begin{eqnarray}
  \label{eq:Hamiltonian0}
  H =\Delta \sigma_{22} 
  +(\Omega \sigma_{21}+g a \sigma_{20}+ {\rm{H.c.}}) \, ,
\end{eqnarray}
and it can be easily diagonalized in the basis being the product
states of the atomic states and the cavity field photon states. To
give the expressions a more compact form, we denote by
$|jn\rangle=|j\rangle\otimes |n\rangle$ a state of the system
consisting of the atomic state $|j\rangle$ and the cavity field with
$n$ photons. Diagonalizing Hamiltonian~(\ref{eq:Hamiltonian0}) in the
basis $\{|1\,n\rangle,|0\,n+1\rangle,|2\,n\rangle\}$ leads to the
dressed states energies~\cite{parkins93:_synth_zeeman}
\begin{eqnarray}
  \label{eq:1}
  \omega_{0}=0,&\quad& \omega_{\pm}=\frac{1}{2}(\Delta\pm\Omega')\, ,
\end{eqnarray}
where
\begin{eqnarray}
  \label{eq:2}
  \Omega'=\sqrt{\Delta^{2}+4g^{2}(n+1)+4\Omega^{2}},
\end{eqnarray}
and the dressed states
\begin{eqnarray}
  \label{eq:3}
  |\Psi_{0}\rangle&=&-\sin\vartheta|1\,n\rangle
  +\cos\vartheta|0\,n+1\rangle\nonumber\\
  |\Psi_{-}\rangle&=&-\sin\varphi\left[\cos\vartheta|1\,n\rangle
    +\sin\vartheta|0\,n+1\rangle\right]\nonumber\\
  &&+\cos\varphi|2\,n\rangle\\ 
  |\Psi_{+}\rangle&=&\cos\varphi\left[\cos\vartheta|1\,n\rangle
    +\sin\vartheta|0\,n+1\rangle\right]\nonumber\\
  &&+\sin\varphi|2\,n\rangle\, .\nonumber\ 
\end{eqnarray}
We have introduced the notation
\begin{eqnarray}
  \label{eq:4}
  \sin\vartheta=\frac{\tilde{g}}{\sqrt{1+\tilde{g}^{2}}},&\quad&
  \cos\vartheta=\frac{1}{\sqrt{1+\tilde{g}^{2}}}\, ,\\
  \sin\varphi=\sqrt{\frac{\Omega'+\Delta}{2\Omega'}},&\quad&
  \cos\varphi=\sqrt{\frac{\Omega'-\Delta}{2\Omega'}}\, ,
\end{eqnarray}
where $\tilde{g}=g\sqrt{n+1}/\Omega$.

Initially, the cavity field mode is in a vacuum state and, therefore,
we are especially interested in the time evolution of the $|10\rangle$
state, so we assume $n=0$. Knowing the evolution for the dressed
states~(\ref{eq:3}), we can write the exact expression for this
evolution of the initial state $|10\rangle(t)$
\begin{equation}
  \label{eq:evL}
  \textrm{e}^{-i H t}|10\rangle =|10\rangle(t)=a(t)|10\rangle
  +b(t)|01\rangle +c(t)|20\rangle,
\end{equation}
where
\begin{eqnarray}
  \label{eq:5}
  a(t)&=&\cos^{2}\vartheta\left[\tilde{g}^{2}+f_{+}(t)
    -\frac{\Delta}{\Omega'}f_{-}(t)\right],\nonumber\\  
  b(t)&=&\sin\vartheta\cos\vartheta\left[-1+f_{+}(t)
    -\frac{\Delta}{\Omega'}f_{-}(t)\right],\\ 
  c(t)&=&\frac{2\Omega}{\Omega'}f_{-}(t),\nonumber
\end{eqnarray}
and
\begin{eqnarray}
  \label{eq:6}
  f_{\pm}(t)&=&\frac{1}{2}\left(\textrm{e}^{-i\omega_{+}t}
    \pm\textrm{e}^{-i\omega_{-}t}\right)\, .
\end{eqnarray}
The perfect $\pi$ pulse operation, defined by
$|10\rangle\to|01\rangle$, requires the condition $\tilde{g}=1$, which
means $g=\Omega$ and, therefore, we restrict the following
considerations to this case only.  Then, we have
$\cos\vartheta=\sin\vartheta=1/\sqrt{2}$, and expressions~(\ref{eq:5})
take simpler form.

It is evident from~(\ref{eq:5}) that the excited atomic state
population shows oscillatory behavior.  Since the qubit is coded into
the lower states, the fidelity of quantum operations performed on the
qubit is reduced when some amount of population is present in the
intermediate level. To overcome this problem, the upper level is
usually adiabatically eliminated by choosing the detuning so large as
to make the intermediate state population negligible. There is,
however, an important drawback of this approach --- it is necessary to
use very large values of the detuning to achieve fidelities of the
quantum operations sufficient for quantum computation. It would
require $\Delta/g>640$ to get the fidelity different from unity by
amount smaller than $10^{-5}$. Such detunings lead to long operation
times, which are very challenging for optical cavities.

Here, we propose a scheme for fine tuning the quantum operations by
using the fact that, in an ideal case, the population of the excited
state evolves periodically in time, periodically approaching zero. By
quantum operation, we understand a unitary operation transforming a
given initial state into another (desired) quantum state. The quantum
operation is perfect if the desired state is produced with the
fidelity equal to unity. 
The idea of fine tuning is to choose the operation time in a way as to
complete the operation when the population of the intermediate state
is zero. To achieve this goal, we choose the evolution time in such a
way that $f_{-}(t=t_{k})=0$, which means
\begin{eqnarray}
  \label{eq:7}
  (\omega_{+}-\omega_{-})\,t_{k}=\Omega'\,t_{k}=2\pi\, k,&\quad& k=1,2,3\ldots,
\end{eqnarray}
and
\begin{eqnarray}
  \label{eq:8}
  f_{+}(t_{k})=\exp\left[i\frac{\epsilon}{2}(\Omega'-|\Delta|)t_{k}\right]
\end{eqnarray}
with
\begin{eqnarray}
  \label{eq:9}
  \epsilon=\left\{
    \begin{array}{rl}
      1&\textrm{ if }\Delta\ge 0\\
      -1&\textrm{ if }\Delta <0
    \end{array}
  \right. \, .
\end{eqnarray}

\begin{figure}[htbp]
  \centering
  \includegraphics[width=7cm]{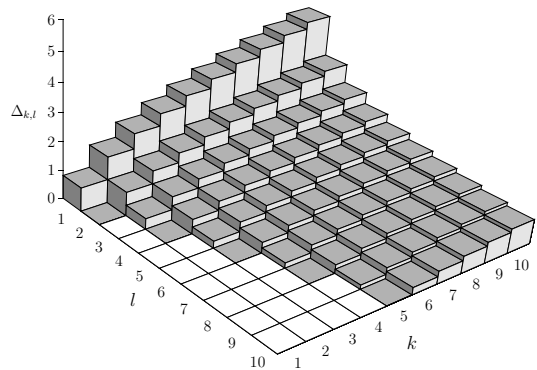}
  \caption{Discrete values of $|\Delta_{k,l}|$ (in units of 2$g$) for
    initial values of $k$ and $l$.}
  \label{fig:1}
\end{figure}

Let us now require that, beside the relation~(\ref{eq:7}), the
evolution time satisfies the relation
\begin{eqnarray}
  \label{eq:10}
  \frac{1}{2}(\Omega'-|\Delta|)\,t_{k,l}=\frac{\pi}{2}\,l,\quad l=1,2,3,\ldots,2k-1 \, .
\end{eqnarray}
Both requirements for the evolution time can be satisfied provided the
numbers $k$ and $l$ obey the relation
\begin{eqnarray}
  \label{eq:11}
  \frac{2k}{l}=\frac{\Omega'}{\Omega'-|\Delta|}\ge 1,&\quad& 2k\ge l,
\end{eqnarray}
which leads to the discrete set of detunings $\Delta_{k,l}$ given by
the formula
\begin{eqnarray}
  \label{eq:12}
  \left(\frac{\Delta_{k,l}}{2g}\right)^{2}
  =\frac{2\left(\frac{2k}{l}-1\right)^{2}}{2\left(\frac{2k}{l}-1\right)+1}\, .
\end{eqnarray}
For large detunings $2k/l-1$ is large, and we can drop unity in the
denominator getting simpler, but approximate, relation. The discrete
values of the detuning $\Delta_{k,l}$, in units of $2g$, are shown in
Fig.~\ref{fig:1}

Enforcing both condition~(\ref{eq:7}) and~(\ref{eq:10}) gives us a
discrete set of the evolution times for which solution~(\ref{eq:evL})
takes the form
\begin{eqnarray}
  \label{eq:13}
  |10\rangle(t_{k,l})=\frac{1}{2}\left[1+(i\epsilon)^{l}\right]|10\rangle
  -\frac{1}{2}\left[1-(i\epsilon)^{l}\right]|01\rangle \, .
\end{eqnarray}
The solution~(\ref{eq:13}) is quite simple, and under appropriate
choice of the numbers $\{k,l\}$ it gives either the superposition of
the two initial states or one of the initial states.

The important $\pi$ operation we get for times
\begin{eqnarray}
  \label{eq:pipulse}
  t_{\pi}=t_{k,l}, &\quad& \textrm{ for }l\textrm{ even, }\quad l/2
  \textrm{ odd }  
\end{eqnarray}
for which
\begin{eqnarray}
  \label{eq:perfPi}
  \textrm{e}^{-i H t_{\pi}}|10\rangle &=&-|01\rangle \, .
\end{eqnarray}

For the second basic operation, the $\pi/2$ pulse operation, we
should choose $l$ odd, for which we have
\begin{eqnarray}
  \label{eq:pi2pulse}
  t_{\pi/2}=t_{k,l}, &\quad& \textrm{ for }l\textrm{ odd, }
\end{eqnarray}
and the solutions are
\begin{equation}
  \label{eq:perfPi2}
  \textrm{e}^{-i H t_{\pi/2}}|10\rangle =\left\{
    \begin{array}{ll}
      \frac{1-i\epsilon}{2}|10\rangle
      -\frac{1+i\epsilon}{2}|01\rangle,&
      \frac{l-1}{2}\textrm{ odd }\\[6pt] 
      \frac{1+i\epsilon}{2}|10\rangle-\frac{1-i\epsilon}{2}|01\rangle,&
      \frac{l-1}{2}\textrm{ even }
    \end{array}\right. \, .
\end{equation}
Depending on the value of $(l-1)/2$, we get from~(\ref{eq:perfPi2}) one
of the two orthogonal superposition states, which are maximally
entangled (Bell states) of the atom-cavity system.

The solutions presented above are a direct consequence of the periodic
evolution of the system. The periodicity appears when the two
frequencies $\omega_{\pm}$, given by~(\ref{eq:1}), are commensurate,
i.e., their ratio is a ratio of integers. Taking absolute values of
the two frequencies we can distinguish between the ``fast''
$\omega_{>}$ and ``slow'' $\omega_{<}$ frequency, where
\begin{eqnarray}
  \label{eq:14}
  \omega_{>\atop <}=\frac{1}{2}\left(\Omega'\pm|\Delta_{k,l}|\right)=\left\{
    \begin{array}{ll}
      |\omega_{\pm}|&\textrm{ if } \Delta_{k,l} >0\\
      |\omega_{\mp}|&\textrm{ if } \Delta_{k,l} < 0
    \end{array}\right. \, .
\end{eqnarray}
From~(\ref{eq:7}) and~(\ref{eq:10}) it is easy to check that the
frequencies $\omega_{<}$, $\omega_{>}$, and
$\Omega'=\omega_{<}+\omega_{>}$ are all commensurate, so the evolution
is periodic.

The period of the slow oscillation is equal to
\begin{eqnarray}
  \label{eq:15}
  T=\frac{2\pi}{\omega_{<}}=\frac{4\pi}{\Omega'-|\Delta_{k,l}|}
  =4\,\frac{t_{k,l}}{l}\, ,
\end{eqnarray}
and is related to the time $t_{k,l}$ for perfect quantum operation. The
period of oscillation of the intermediate state population is given by
\begin{eqnarray}
  \label{eq:16}
  T'=\frac{2\pi}{\Omega'}=\frac{2\pi}{\omega_{<}}\frac{\omega_{<}}{\Omega'}
  =T\frac{l}{4k}=\frac{t_{k,l}}{k} \, .
\end{eqnarray}
The subsequent minima in the intermediate state population are
separated by $T'$. If $4k/l$ is an integer the period of the system
evolution is equal to $T$, and, simultaneously, it is equal to an
integer multiple ($4k/l$) of the period $T'$. When $4k/l$ is an
irreducible fraction, the period of the system evolution is equal to
$4kT'=lT$, and if the fraction is reducible the period is reduced
appropriately.

For example, choosing $k=1$ and $l=2$, we have $\Delta_{k,l}=0$ and
the resulting state is, according to~(\ref{eq:perfPi}), $-|01\rangle$,
which is, up to the phase, an illustration of the perfect $\pi$ pulse
operation. The 
operation is completed at time $t_{k,l}=k T'$ for which the population
of the intermediate state is zero, so the fidelity of the operation is
equal to one.  Similarly, by choosing $k=1$ and $l=1$ we
get, according to~(\ref{eq:perfPi2}), the superposition state
$(1+i)/2|01\rangle-(1-i)/2|10\rangle$ which is generated with the
fidelity equal to one.  This is a perfect $\pi/2$ pulse operation, and
the detuning in this case is equal to $\Delta_{1,1}/(2g)=0.8165$.

The two examples illustrate the idea of fine tuning of the quantum
operations: choose one of the discrete values of the detuning
$\Delta_{k,l}$ and corresponding operation time $t_{k,l}$ to complete
the operation at a time when the population of the intermediate state is
zero.

\section{Approximate solutions}
So far we have discussed the ideal case, when there is no cavity
decay, and the spontaneous emission from the atomic excited level is
ignored.  To make the system useful for quantum information processing
it is necessary to have access to the quantum information stored in
the system and the possibility of sending it over long distances. In the case
of the quantum system under discussion it is possible, when one mirror
of the cavity is partially transparent. Of course, transparency of the
mirror leads to a damping of the cavity field mode.  Moreover,
spontaneous emission introduces damping to the atomic system which
spoils the desirable results of quantum operations.  Unfortunately, in the presence of
damping in the system, it is not possible to get exact analytical
solutions; therefore, we have to resort to some approximations.  When
the two decay rates $\kappa$ and $\gamma$ are small we can apply the
perturbative methods to find the first order corrections to the
dressed states energies as well as the state amplitudes.

The first order corrections to the dressed states energies lead to the
following modifications:
\begin{eqnarray}
  \label{eq:17}
  \tilde{\omega}_{0}=\omega_{0}-i\kappa_{0},&\quad&
  \tilde{\omega}_{\pm}=\omega_{\pm}-i(\kappa_{\pm}+\gamma_{\pm}) \, ,
\end{eqnarray}
where the damping rates $\kappa_{0},\kappa_{\pm}$, and $\gamma_{\pm}$
are given by
\begin{equation}
  \label{eq:18}
  \kappa_{0}=\frac{\kappa}{2},\quad
  \kappa_{\pm}=\frac{\kappa}{4}\left(1\mp d\right),\quad
  \gamma_{\pm}=\frac{\gamma}{2}(1\pm d) \, ,
\end{equation}
where $d=\Delta/\Omega'$.  The solution~(\ref{eq:evL}) for the
conditional evolution of the state $|10\rangle$, according to the
first order perturbation theory with respect to both $\kappa$ and
$\gamma$, is given, under the assumption $\Omega=g$ ($\tilde{g}=1$),
by the following formulas:
\begin{eqnarray}
  \label{eq:19}
  a(t)&=&\frac{1}{2}\left\{(1+i\,2\eta\, d)\,\textrm{e}^{-\kappa_{0}t}
    +(1-i\,2\eta\, d)\,\tilde{f}_{+}(t)\right.\nonumber\\
  &&\left.-\left[d-i\,\xi-i\,\eta\,(1+d^2)\right]\tilde{f}_{-}(t)\right\}\nonumber\\
  b(t)&=&\frac{1}{2}\left\{-\textrm{e}^{-\kappa_{0}t}
    +\tilde{f}_{+}(t)
    -\left(d-i\,\xi\right)\,\tilde{f}_{-}(t)\right\}\nonumber\\
  c(t)&=&\frac{1}{\Theta}\left\{-i\,\frac{\eta}{2}\textrm{e}^{-\kappa_{0} t}
    +i\,\frac{\eta}{4}(1+d^2)\,\tilde{f}_{+}(t)\right.\nonumber\\
  &&\left.+\left[1-i\,\left(\frac{\eta}{2}-\bar{\gamma}\right)\, 
  d\right]\tilde{f}_{-}(t)\right\}\, ,
\end{eqnarray}
where we have introduced the notation $\Theta=\Omega'/(2g)$,
$\bar{\gamma}=\gamma/\Omega'$, $\bar{\kappa}=\kappa/\Omega'$,
$\eta=\bar{\kappa}\Theta^{2}$, and $\xi=\bar{\gamma}/\Theta^{2}$.
Functions $\tilde{f}_{\pm}$ have the form~(\ref{eq:6}) except that the
frequencies $\omega_{\pm}$ are replaced by $\tilde{\omega}_{\pm}$
from~(\ref{eq:17}).

The solution~(\ref{eq:19}) is valid as long as $\kappa$ and $\gamma$
are small. In fact, the real smallness parameters are $\eta$ and $\xi$,
so we require that both $\eta\ll 1$ and $\xi\ll 1$. In
deriving~(\ref{eq:19}) we also discarded terms proportional to the
product $\kappa\gamma$.  On the other hand, the solution~(\ref{eq:19})
is valid for any value of the detuning $\Delta$. This means that it
allows also for the resonant case of the $\pi$ pulse operation, when
$k=1,l=2$ and $\Delta_{1,2}=0$.  The resonant case is usually not
recommended because of the spontaneous emission from the intermediate
level. What is usually done to minimize the effect of spontaneous
emission is the adiabatic elimination of the excited level. This
requires large values of the detuning $\Delta$. It is seen
from~(\ref{eq:19}) that when $1/\Theta=2g/\Omega'\ll 1$, the amplitude
$c(t)$ of the excited level is small. Neglecting this amplitude by
setting $c(t)=0$ is exactly what the adiabatic elimination is about.

However, when the rate of spontaneous emission is small with respect
to $2g$, the approximate solution~(\ref{eq:19}) should describe
properly the role of spontaneous emission. This will be discussed
later.

\subsection{Adiabatic elimination}
The standard procedure used to eliminate the influence of the exited
atomic level on the result of quantum operation is the adiabatic elimination of the
excited level. It is realized by taking large detunings of the fields
from the atomic transition frequencies, which makes the population of
the excited state very small, and, in consequence, the state is
assumed not to take part in the evolution. Assuming that
$1/\Theta=2g/\Omega'\ll 1$, which means that $\Delta/(2g)\gg 1$ and
the requirements for adiabatic elimination are met, we can eliminate
the state $|20\rangle$ from the evolution by putting in
equations~(\ref{eq:19}) $c(t)=0$. We should also put $\gamma=0$
($\bar{\gamma}=0,\xi=0$) to get rid of the $\gamma$ dependence in the
other amplitudes. For large detuning we can also put $|d|=1$. With all
these substitutions we get the following solutions:
\begin{eqnarray}
  \label{eq:20}
  a(t)&=&\frac{1}{2}\,\textrm{e}^{-\kappa_{0}t}\left[1+2i\epsilon\eta+(1-2i\epsilon\eta)
    \textrm{e}^{i\epsilon\omega_{<}t}\right] \nonumber\\
  &=&\textrm{e}^{-\kappa_{0}t}\,\textrm{e}^{\frac{i}{2}\epsilon\omega_{<}t}
  \left[\cos \left(\frac{1}{2}\omega_{<}t\right)
    +2\eta\sin \left(\frac{1}{2}\omega_{<}t\right)\right]\nonumber\\
  b(t)&=&\frac{1}{2}\,\textrm{e}^{-\kappa_{0}t}\left(-1+
    \textrm{e}^{i\epsilon\omega_{<}t}\right)\nonumber\\
  &=&\textrm{e}^{-\kappa_{0}t}\,\textrm{e}^{\frac{i}{2}\epsilon\omega_{<}t}
  \left[i\epsilon\sin\left(\frac{1}{2}\omega_{<}t\right)\right]\, ,
\end{eqnarray}
where now $\eta=\kappa|\Delta|/(4g^{2})$. From~(\ref{eq:20}), one can
calculate the time for the $\pi$ pulse operation, which takes the form
\begin{eqnarray}
  \label{eq:21}
  t_{\pi}(\kappa)=t_{k,l}\left[1+\frac{2}{l}\left(1
      -\frac{2}{\pi}\arctan\frac{1}{2\eta}\right)\right]\, ,
\end{eqnarray}
and similarly for the $\pi/2$ pulse operation
\begin{eqnarray}
  \label{eq:22}
  t_{\pi/2}(\kappa)=t_{k,l}\left[1
    -\frac{1}{l}\left(1-\frac{4}{\pi}\arctan
      \frac{1}{1-2\eta}\right)\right]\, .
\end{eqnarray}
For large detuning $\omega_{<}=2g^{2}/|\Delta|$,
$2\eta=\kappa/\omega_{<}$, and formulas~(\ref{eq:21})
and~(\ref{eq:22}) are consistent with corresponding formulas obtained
when the adiabatic elimination is performed at the Hamiltonian
level~\cite{bose}.  To be precise, the consistency is up to terms
linear in $\kappa$, because the solutions~(\ref{eq:19}), and
consequently~(\ref{eq:20}), were obtained in the linear approximation
with respect to $\kappa$ (or $\eta$), i.e., under assumption $\eta\ll
1$.  As far as the linear approximation is valid, one can derive
simplified formulas for the relative changes of the operation
times, which take the form
\begin{eqnarray}
  \label{eq:23}
  \frac{t_{\pi}(\kappa)-t_{k,l}}{t_{k,l}}&=&\frac{8\eta}{l\pi}
  \nonumber\\
  \frac{t_{\pi/2}(\kappa)-t_{k,l}}{t_{k,l}}&=&\frac{4\eta}{l\pi}\, .
\end{eqnarray}

In the presence of damping the evolution is not unitary and the
resulting state, under the condition that neither a photon from the
cavity nor a spontaneous emission photon are registered, is given by
\begin{equation}
  \label{eq:24}
  |\Psi(t)\rangle=\frac{|\tilde{\Psi}(t)\rangle}{\sqrt{\langle\tilde{\Psi}(t)
      |\tilde{\Psi}(t)\rangle}}\, ,
\end{equation}
where
\begin{equation}
  \label{eq:25}
  |\tilde{\Psi}|(t)\rangle=\text{e}^{-iHt}|10\rangle
\end{equation}
is the unnormalized, conditional quantum state generated with the
effective Hamiltonian~(\ref{eq:Hamiltonian0L}). The fidelity of the
resulting state is equal to $|\langle\Psi(t)|10\rangle(t_{k,l})|^2$,
where $|10\rangle(t_{k,l})$ is a desired state given
by~(\ref{eq:13}). The solutions~(\ref{eq:20}) are the amplitudes of
the unnormalized state~(\ref{eq:25}) and must be normalized when we
calculate state~(\ref{eq:24}) and the fidelity in the presence of
damping. For simplicity we use 
the same notation for the normalized amplitudes $a,b,c$ as for the
unnormalized.

It is well known that when damping is present in the system, the
periodic behavior of the system is lost, and the maxima and minima are
shifted. This is exactly what we observe here. The quantum states
generated in the presence of damping are no longer the ideal
states~(\ref{eq:perfPi}) or~(\ref{eq:perfPi2}), but if $\kappa/(2g)\ll
1$, the fidelity of the states generated can be quite high. The best
choice for $t_{k,l}$ is that with the smallest possible values of $l$.
\begin{figure}[h]
  \centering
  \includegraphics[width=7cm]{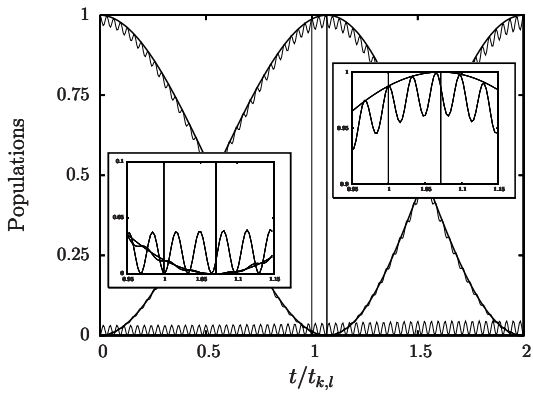}
  \caption{Comparison of the populations $|a(t)|^{2}$, $|b(t)|^{2}$,
    and $|c(t)|^{2}$ obtained from Eqs.~(\ref{eq:19})
    and~(\ref{eq:20}) (normalized) for $k=31$, $l=2$, and
    $\kappa/(2g)=0.01$. Broad lines are populations after adiabatic
    elimination of the intermediate level. Vertical lines mark the
    times $t_{k,l}$ and $t_{\pi}(\kappa)$.}
  \label{fig:2}
\end{figure}
An example is shown in Fig.~\ref{fig:2}, where the $\pi$ pulse
operation is illustrated. After adiabatic elimination of the
intermediate level the populations of the remaining levels oscillate
smoothly (broad lines), and the maximum is shifted from the time
$t_{k,l}$ to $t_{\pi}(\kappa)$. However, without adiabatic elimination
the populations are modulated with fast oscillations, and there are
fast oscillations of the population of the intermediate level seen at
the bottom. With the parameters of Fig.~\ref{fig:2} the detuning
$\Delta_{31,2}/(2g)=5.4321$. If $g/2\pi=16$ MHz then
$\Delta_{31,2}/2\pi=173.83$~MHz, which is quite big, but the presence
of the intermediate level is still visible, and it has significant
influence on the fidelity of the resulting state. In the figures we
illustrate the state evolution by plotting the state populations only,
but the solutions are more general and give the state amplitudes that
include also the phase information. The resulting state is a
superposition of the basis states and the populations do not fully
characterize the superposition. More precise characteristic of the
state created during the evolution is its fidelity. To make it clear
we give examples of the state amplitudes and the fidelities in
Tables~\ref{tab:1} and~\ref{tab:2}. For the example shown in
Fig.~\ref{fig:2}, we present in Table~\ref{tab:1} the amplitudes,
calculated according to the approximate solution~(\ref{eq:19}), for
the state amplitudes at different operation times.
\begin{table}[h]
  \caption{Normalized amplitudes of the quantum states for the $\pi$ pulse operation at
    various operation times ($k=31$, $l=2$, $\kappa/(2g)=0.01$)}
\label{tab:1} 
{\scriptsize
\begin{tabular}[t]{|c|c|c|c|c|}\hline\hline
  Time&a&b&c&Fidelity\\\hline
  $t_{k,l}$&-0.0029+0.1078\,i&-0.9940+0.0000\,i&-0.0162-0.0098\,i&0.9880\\\hline
  $t_{\pi}(\kappa)$&0.0022+0.0066\,i&-0.9887-0.1012\,i&0.0372+0.1043,i&0.9877\\\hline
  $t_{f}$&-0.0037+0.0064\,i&-0.9946-0.1015\,i&-0.0162-0.0133\,i&0.9995\\\hline\hline
\end{tabular}
}
\end{table}

\begin{figure}[h]
  \centering
  \includegraphics[width=7cm]{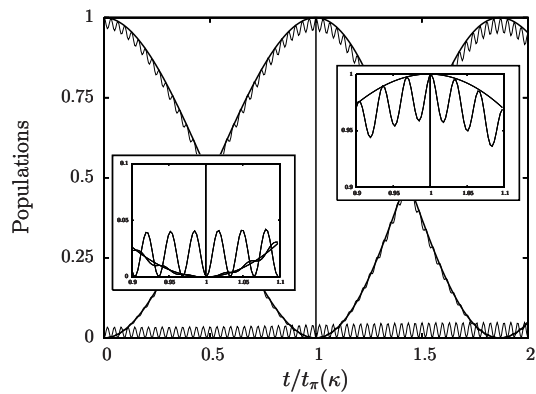}
  \caption{Same as Fig.~\ref{fig:2}, but with $\Delta_{k,l}(\kappa)$
    given by~(\ref{eq:26}). Vertical line marks the
    $t_{\pi}(\kappa)$.}
  \label{fig:3}
\end{figure}

The time $t_{f}$, which we denote as the fine tuning time, is defined
as the time at which the closest to $t_{\pi}(\kappa)$ maximum of the
population of the state $|01\rangle$ (or minimum of the population of
the state $|20\rangle$) occurs. From insets in Fig.~\ref{fig:2} it is
clear that $t_{f}=t_{k,l}+2T'$, and this choice improves significantly
the fidelity of the operation. Similar results are obtained for the
$\pi/2$ pulse operation. An example is given in Table~II, where the
state amplitudes are presented for $k=31$, $l=1$, and
$\kappa/(2g)=0.01$. In this case $\Delta_{31,1}/(2g)=7.7784$, and it
is larger than in the previous case. This example is even more
interesting because the fidelity at the operation time
$t_{\pi/2}(\kappa)$ coming from the adiabatic elimination is smaller
than the fidelity at time $t_{k,l}$ for ideal case of periodic
evolution. The fine tuning, which this time gives $t_{f}=t_{k,l}+3T'$,
again improves the fidelity.
\begin{table}[h]
  \caption{Normalized amplitudes of the quantum states for the $\pi/2$ pulse operation at
    various operation times ($k=31$, $l=1$, $\kappa/(2g)=0.01$)}
\label{tab:2} 
{\scriptsize
\begin{tabular}[t]{|c|c|c|c|c|}\hline\hline
  Time&a&b&c&Fidelity\\\hline
  $t_{k,l}$&0.5342+0.5351\,i&-0.4623+0.4632\,i&-0.0045+0.0029\,i&0.9948 \\\hline
  $t_{\pi/2}(\kappa)$&0.4584+0.5339\,i&-0.5371+0.4497\,i&0.0589-0.1034\,i&0.9858\\\hline
  $t_{f}$&0.4632+0.5407\,i&-0.5323+0.4579\,i&-0.0058+0.0029\,i&0.9999\\\hline\hline
\end{tabular}
}
\end{table}
From Fig.~\ref{fig:2} it is clear that the time $t_{\pi}(\kappa)$ is
different from the time $t_{k,l}$ for the ideal case. This difference
suggests another possibility of fine tuning of the quantum
operations, which consists in taking the operation time calculated
from~(\ref{eq:21}) as a new $t_{k,l}$ and adjust the detuning
appropriately to get a maximum of the fast oscillations at the new
$t_{k,l}$. Within linear approximation with respect to $\eta$ it leads
to the detuning
\begin{eqnarray}
  \label{eq:26}
  |\Delta_{k,l}(\kappa)|=|\Delta_{k,l}|\left[1-\frac{4\eta}{l\pi}
    \sin^{2}\left(\frac{l\pi}{4}\right)\right].
\end{eqnarray}
By adjusting the detuning according to~(\ref{eq:26}) we get the
situation illustrated in Fig.~\ref{fig:3}.
Now, the fidelity at time $t_{\pi}(\kappa)$ takes the value $0.9992$.
The fine tuning of the quantum operations can be performed in
both ways: by finding the best operation time and/or adjusting the
detuning. The values $\Delta_{k,l}$ given by~(\ref{eq:12}) for the ideal
case are a good starting point.

Examples given here show clearly that even for quite large values of
detuning, the presence of the intermediate state is still visible, but
the fidelity of the resulting states can be significantly enhanced when the
operation time and/or the detuning are adjusted appropriately. This,
however, requires knowledge of the solutions for the three-level
system, and cannot be done when the adiabatic elimination of the
intermediate state has already been accomplished.

Making the adiabatic elimination on the Hamiltonian level, consisting in
diagonalizing the $2\times 2$ resulting Hamiltonian without resorting
to perturbation theory, leads to the following
results~\cite{bose}:
\begin{eqnarray}
  \label{eq:27}
  a(t)&=&\textrm{e}^{\frac{i}{2}\left(\epsilon\omega_{<}+i\kappa\right)t}
  \left[\cos\left(\frac{1}{2}\,\omega_{<}(\kappa)
      t\right)\right.\nonumber\\
  &&\left.\hphantom{\textrm{e}^{\frac{i}{2}\left(\epsilon\omega_{<}+i\kappa\right)t}}
    +\frac{\kappa}{\omega_{<}(\kappa)}
    \sin\left(\frac{1}{2}\,\omega_{<}(\kappa)t\right)\right],\nonumber\\
  b(t)&=&\textrm{e}^{\frac{i}{2}\left(\epsilon\omega_{<}+i\kappa\right)t}
  \left[i\epsilon\frac{\omega_{<}}{\omega_{<}(\kappa)}
    \sin\left(\frac{1}{2}\,\omega_{<}(\kappa)t\right)\right],
\end{eqnarray}
where $\omega_{<}(\kappa)=\sqrt{\omega_{<}^{2}-\kappa^{2}}$ is the
modified by $\kappa$ frequency $\omega_{<}$. Since the modification is
at least quadratic in $\kappa$, by keeping only the linear terms
in~(\ref{eq:27}), we easily reproduce formulas~(\ref{eq:20}). When
corrections due to $\kappa^{2}$ become important, the populations of
the two states oscillate with the modified frequency
$\omega_{<}(\kappa)$ instead of $\omega_{<}$. The operation
times are then given by
\begin{eqnarray}
  \label{eq:28}
  t_{\pi}(\kappa)&=&\frac{t_{k,l}}{r}\left[1+\frac{2}{l}
    \left(1-\frac{2}{\pi}\arctan\frac{r}{\sqrt{1-r^2}}\right)\right],\nonumber\\
  t_{\pi/2}&=&\frac{t_{k,l}}{r}\left[1-\frac{1}{l}
    \left(1-\frac{4}{\pi}\arctan\frac{r}{1-\sqrt{1-r^2}}\right)\right]\,\hspace*{12pt}
\end{eqnarray}
where $r$ is the ratio given by
\begin{eqnarray}
  \label{eq:29}
  r&=&\omega_{<}(\kappa)/\omega_{<}=\sqrt{1-(\kappa/\omega_{<})^{2}}.
\end{eqnarray}
Since $\omega_{<}=2g^{2}/|\Delta|$ decreases as $|\Delta|$ increases,
the ratio $r$ for a given value of $\kappa$ can significantly differ
from unity. This is true for very large detunings, i.e., for large
values of $k$. It is important, however, to have $\kappa/\omega_{<}<
1$, which puts some restrictions on the possible values of the
detuning.

\subsection{Small detunings and spontaneous emission}
In the case of small detunings, adiabatic elimination of the intermediate
state is not possible, but solutions~(\ref{eq:19}) are valid for
arbitrary detuning, that is, also for small detunings and resonant
cases.  In order to check how close to unity the fidelity of the
$\pi$ pulse operation can be, we have investigated this problem for the
parameters $(\Omega,g,\kappa)/2\pi=(16,16,0.05)$ MHz. This means that
$\kappa/(2g)=0.0015$. Of course, for nonzero cavity damping $\kappa$,
we should choose the smallest possible values for $l$ as to complete
the operation in the shortest possible time.  For example, for $k=1$,
$l=2$ we get from~(\ref{eq:12}) $\Delta_{k,l}=0$ and we obtain, after
time $t_{k,l}$, the state $|01\rangle$ with the fidelity different
from unity by less than $10^{-6}$.  This is the resonant case, which
is usually not recommended because of the spontaneous emission from
the excited atomic state.

Similarly for the $\pi/2$ pulse operation, for $k=1$, $l=1$ we get
$\Delta/(2g)=0.8165$, which means that for $g/2\pi=16$~MHz,
$\Delta/2\pi=26.128$~MHz, and for such a value of $\Delta$ we can
reproduce the state~(\ref{eq:perfPi2}) with the fidelity that differs
from unity by less than $10^{-6}$.

Choosing $k=4$ and $l=2$, we get the $\pi$ operation with
$\Delta_{k,l}/(2g)=1.6036$ ($51.314$ MHz for $g/2\pi=16$~MHz), and we
have a noticeable value of detuning. For $\kappa/2\pi=0.05$ MHz, we
still get fidelity $0.99997$.

\begin{figure}[h]
  \centering
  \includegraphics[width=7cm]{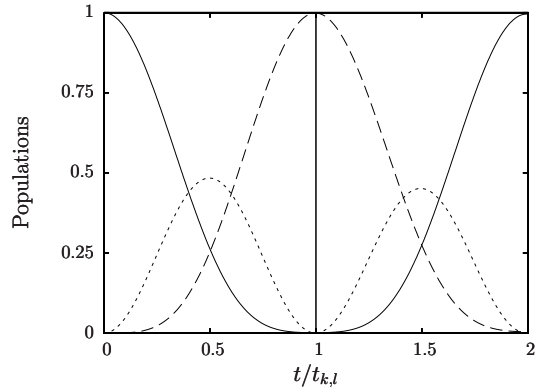}
  \caption{Evolution of the atom-field states populations
    (normalized): $|a(t)|^{2}$ (solid line), $|b(t)|^{2}$ (dashed
    line), $|c(t)|^{2}$ (dotted line) for $k=1$, $l=2$,
    $\kappa/(2g)=0.0015$, and $\gamma/(2g)=0.03$. Time is scaled in
    units of $t_{k,l}$.}
  \label{fig:4}
\end{figure}

If spontaneous emission is ignored, one can expect very good
performance of quantum operations at small detunings. The approximate
formulas~(\ref{eq:19}) take into account spontaneous emission and they
are valid whenever the spontaneous emission rate $\gamma$ is small
with respect to $\Omega'$. We can thus use equations~(\ref{eq:19}) to
calculate the state amplitudes in the presence of spontaneous
emission. Assuming the parameter values
$(\Omega,g,\kappa,\gamma)/2\pi=(16,16,0.05,1)$ MHz, we have
$\kappa/(2g)=0.0015$ and $\gamma/(2g)=0.03$ and, for $k=1$ and $l=2$,
get the situation illustrated in Fig.~\ref{fig:4},
where the state populations are shown. With these values of parameters
the state $|01\rangle$ is generated with the fidelity still equal to
$0.9989$, which is quite good. For the $\pi/2$ pulse operation the
situation is even better and the fidelity takes the value $0.9993$. Of
course, spontaneous emission rapidly deteriorates the quality of
states generated at small detunings, but it is still possible to
obtain remarkable fidelities.

\section{Numerical results}
Approximate analytical results presented in previous section are valid
only for sufficiently small values of $\kappa$ and $\gamma$. They
illustrate the idea of fine tuning of quantum operations, but looking, for
example, at Fig.~\ref{fig:3}, it is clear that it should be feasible
to get even better results when more precise tuning is carried out.
To this end, numerical calculations are necessary.  The effective
Hamiltonian~(\ref{eq:Hamiltonian0L}) can be diagonalized numerically
and we can find nonunitary evolution governed by this Hamiltonian. The
numerical solution allows for adjustment of the detuning and
corresponding time $t_{\pi}$ allowing for further improvement in the
fidelity. For example, for the situation illustrated in
Fig.~\ref{fig:3}, numerical optimization gives the fidelity $0.9997$
if we take $\Delta_{31,2}(\kappa)/(2g)=5.2409$ instead of
$\Delta_{31,2}(\kappa)/(2g)=5.2380$ given by~(\ref{eq:26}) and
$\Delta_{31,2}/(2g)=5.4321$ given by~(\ref{eq:12}), which for
$g/2\pi=16$~MHz gives for the detunings $\Delta_{31,2}/2\pi$ values
$167.71$, $167.62$, and $173.83$~MHz, respectively. This example shows
that for $\kappa/(2g)=0.01$ and $\gamma=0$, analytical formulas give
quite accurate results, and precise numerical tuning does improve the
results but only slightly.

\begin{figure}[h]
  \centering
  \includegraphics[width=7cm]{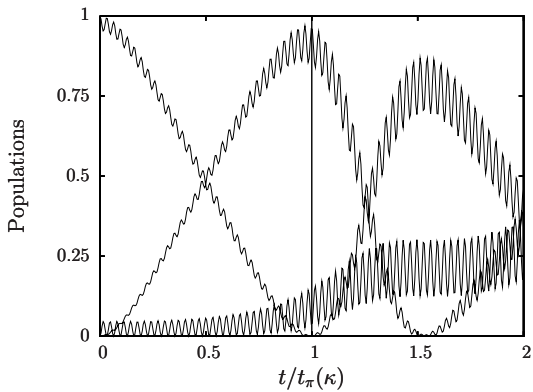}
  \caption{Numerical solutions for the state populations for
    $\kappa/(2g)=0.05$, $\gamma/(2g)=0$, and numerically tuned
    $\Delta_{31,2}/(2g)=4.4491$.}
  \label{fig:5}
\end{figure}

\begin{figure}[h]
  \centering
  \includegraphics[width=7cm]{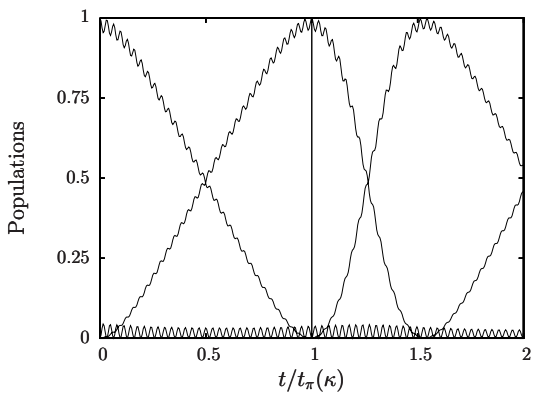}
  \caption{Same as Fig.~\ref{fig:5} but for $\gamma/(2g)=0.03$}
  \label{fig:6}
\end{figure}

The situation changes dramatically when $\kappa$ and $\gamma$ are not
so small as to justify linear approximation made in derivation of the
analytical formulas. In such a situation the only reliable solutions
are numerical solutions. To visualize the role of the two damping
rates in the evolution we plot in Fig.~\ref{fig:5} the populations of
the quantum states for the value of $\kappa/(2g)=0.05$ and $\gamma=0$.
The time $t_{\pi}(\kappa)$ is now calculated according
to~(\ref{eq:28}). The amplitudes of the fast oscillation are quite
big, and the $\pi$ pulse operation is far from being perfect. The
detuning $\Delta_{31,2}$ is numerically tuned to fit the maximum of
population and takes the value $\Delta_{31,2}/(2g)=4.4491$.
We find that the fidelity in this case is equal to $0.9660$ and is, of
course, much worse than it was for $\kappa/(2g)=0.01$ shown in
Fig.~\ref{fig:3}. In both cases the spontaneous emission is not taken
into account. In Fig.~\ref{fig:6} we present the same situation but
when the spontaneous emission is included.
We take the value $\gamma/(2g)=0.03$ for the spontaneous emission
rate, which gives for $g/2\pi=16$~MHz the value $\gamma/2\pi=1$~MHz.
It turns out that the presence of spontaneous emission has a stabilizing
effect on the evolution. The fast oscillations are damped
significantly, and, surprisingly, the fidelity of the state generated
becomes higher. For the parameters of Fig.~\ref{fig:6} we find the
fidelity $0.9994$. The reason for such behavior can be understood from
Eq.~(\ref{eq:18}), which shows the difference between
$\kappa_{\pm}$ and $\gamma_{\pm}$ when the detuning is large and
$|d|\rightarrow 1$. The two damping rates act in a sense in opposite
directions: when fast oscillations are weakly damped by $\kappa$ they
are strongly damped by $\gamma$, and vice versa. The effect is more
pronounced for higher values of the detuning, i.e., for higher values
of $k$. Similar effects can be observed for the $\pi/2$ pulse
operation. These results show that in some situations spontaneous
emission can play a positive role in performance of quantum operations.

\subsection{Nonrectangular pulses}
In real experiments the field cannot be switched on and off abruptly,
so the rectangular pulses are rather not realistic. There are always a
finite rise time and a finite fall time of the pulse. The pulse has
definite shape, and usually what we observe in experiments depends on
the pulse shape. Here, we address the effects of the pulse
shape on the fine tuning discussed above.

With the Hamiltonian~(\ref{eq:Hamiltonian0L}), assuming that
$\Omega\to gF(t)$ and $g\to gF(t)$, where the function $F(t)$
describes the pulse shape, we get 
from the Schr\"odinger equation (units scaled to $2g$) the following
set of equations for the state amplitudes
\begin{eqnarray}
  \label{eq:30}
  \dot{a}(t)&=&-\frac{i}{2}F(t)\,c(t)\, ,\nonumber\\
  \dot{b}(t)&=&-\kappa\, b(t)-\frac{i}{2}F(t)\,c(t)\, ,\\
  \dot{c}(t)&=&-\frac{i}{2}F(t)\,a(t)-\frac{i}{2}F(t)\,b(t)
  -(\gamma+i\Delta)\,c(t)\, .\nonumber
\end{eqnarray}
On resonance, without damping, the system~(\ref{eq:30}) has the
solutions
\begin{eqnarray}
  \label{eq:31}
  a(t)&=&\frac{1}{2}\left[1+f_{+}(t)\right]\, ,\nonumber\\
  b(t)&=&\frac{1}{2}\left[-1+f_{+}(t)\right]\, ,\nonumber\\
  c(t)&=&\frac{1}{\sqrt{2}}\,f_{-}(t)\, .
\end{eqnarray}
where
\begin{eqnarray}
  \label{eq:32}
  f_{\pm}(t)&=&\frac{1}{2}\left\{\exp\left[-\frac{i}{\sqrt{2}}
      \int_{0}^{t}F(t')\text{d}t'\right]\right.\nonumber\\ 
  &&\left.\pm\exp\left[\frac{i}{\sqrt{2}}\int_{0}^{t}F(t')\text{d}t'\right]\right\}\, .
\end{eqnarray}
The solutions~(\ref{eq:31}) depend only on the pulse area
$\int_{0}^{t}F(t')\text{d}t'$, so for any shape $F(t)$, with the same
area, the resulting state after time $t$ is the same.

We do not know analytical solutions for the nonresonant case, even
without damping, but the system~(\ref{eq:30}) can be easily solved
numerically for a given pulse shape $F(t)$. The solutions lack their
periodic character and the quantum operation time depends on the pulse
shape. The situation is thus more complex than for the rectangular
pulses discussed in previous sections. Nevertheless, even in this case
the idea of fine tuning is useful for finding the operation time.
We illustrate the problem with few examples. Let us consider two
different pulse shapes: the trapezium shape and the sine square shape,
which are defined by the functions:
\begin{eqnarray}
  \label{eq:33}
  F(t)=\left\{
    \begin{array}{ll}
      s\frac{t}{t_{r}}&\text{ for } 0<t<t_{r}\\
      s&\text{ for } t_{r}<t<t_{p}-t_{f}\\
      s\frac{t_{p}-t}{t_{f}}& \text{ for } t_{p}-t_{f}<t<t_{p}
    \end{array}\right. \, ,
\end{eqnarray}
where
\begin{equation}
  \label{eq:34}
  s=\frac{1}{1-\frac{t_{r}+t_{f}}{2t_{p}}}
\end{equation}
is the scaling factor which makes the pulse area for the pulse
duration $t_{p}$ to be the same as the area of the rectangular pulse
of the same duration and unit height. The times $t_{r}$ and $t_{f}$
are the rise and fall time of the pulse, respectively. The
normalization to the same pulse area ensures, on resonance, the same
final state at time $t_{p}$ after the pulse for different pulse
shapes.

The sine square pulse shape is given by the function (normalized to
the same area)
\begin{eqnarray}
  \label{eq:35}
  F(t)=\left\{
    \begin{array}{ll}
      2\sin^{2}\left(\pi\frac{t}{t_{p}}\right)&\text{ for } 0<t<t_{p}\\
      0&\text{ otherwise }
    \end{array}\right. \, .
\end{eqnarray}

The two pulse shapes are of different character. The trapezium is
close to the rectangle when the rise and fall times are short. The
sine square pulse is much narrower and changes smoothly over the whole
duration time $t_{p}$. For the rectangle pulse we have $F(t)=1$.

To find approximately the proper operation times for pulse
excitation, we use the following trick. We replace the oscillation
frequency $\Omega'$ of the intermediate state population with an
``average frequency''
\begin{equation}
  \label{eq:36}
  \Omega'=\sqrt{\Delta^{2}+2}\to \Omega'=\sqrt{\Delta^{2}+2\overline{F^{2}(t)}}\, ,
\end{equation}
where
\begin{equation}
  \label{eq:37}
  \overline{F^{2}(t)}=\frac{1}{t_{p}}\int_{0}^{t_{p}}F^{2}(t)\,\text{d}t
\end{equation}
which gives
\begin{equation}
  \label{eq:38}
  \overline{F^{2}(t)}=\left\{
    \begin{array}{ll}
      s(4-s)/3&\text{ for trapezium}\\
      3/2&\text{ for sine square }
    \end{array}\right. \, ,
\end{equation} 
with $s$ given by~(\ref{eq:34}). Applying such a replacement we can
calculate the ``period'' $T'$ using~(\ref{eq:16}) and the time
$t_{k,l}$ for the considered operation using $\Delta_{k,l}$ defined
by~(\ref{eq:12}). The pulse is characterized by one parameter --- the
mean square amplitude defined by~(\ref{eq:37}).  For pulses with
smooth rise and fall times, the evolution is no longer periodic, so
the time $T'$ is not really the period, but for small values of $l$
($l=1,2$), the time $t_{k,l}$ calculated in this way is still a good
approximation for the optimal operation time. It can at least be
treated as a good starting point for further optimization.

\begin{figure}[h]
  \centering
  \includegraphics[width=0.4\textwidth]{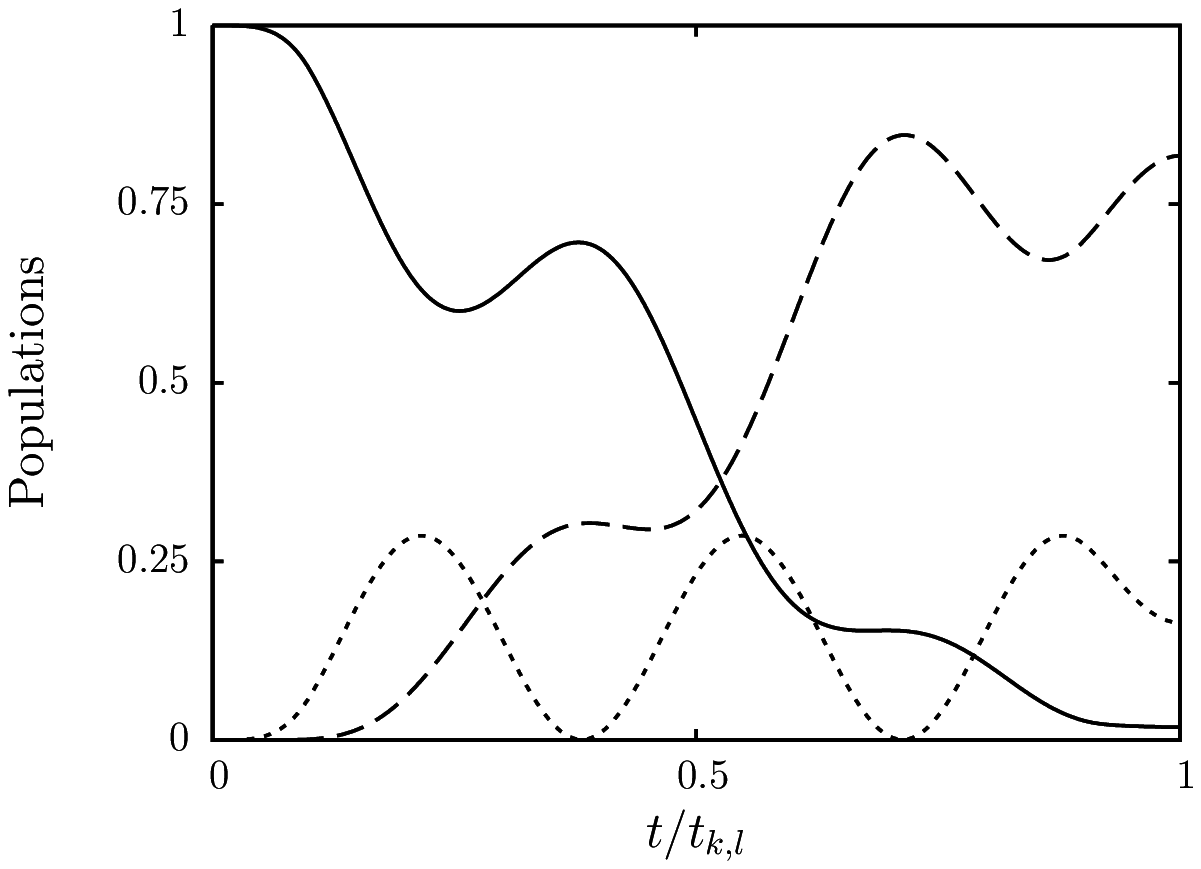}
  \includegraphics[width=0.4\textwidth]{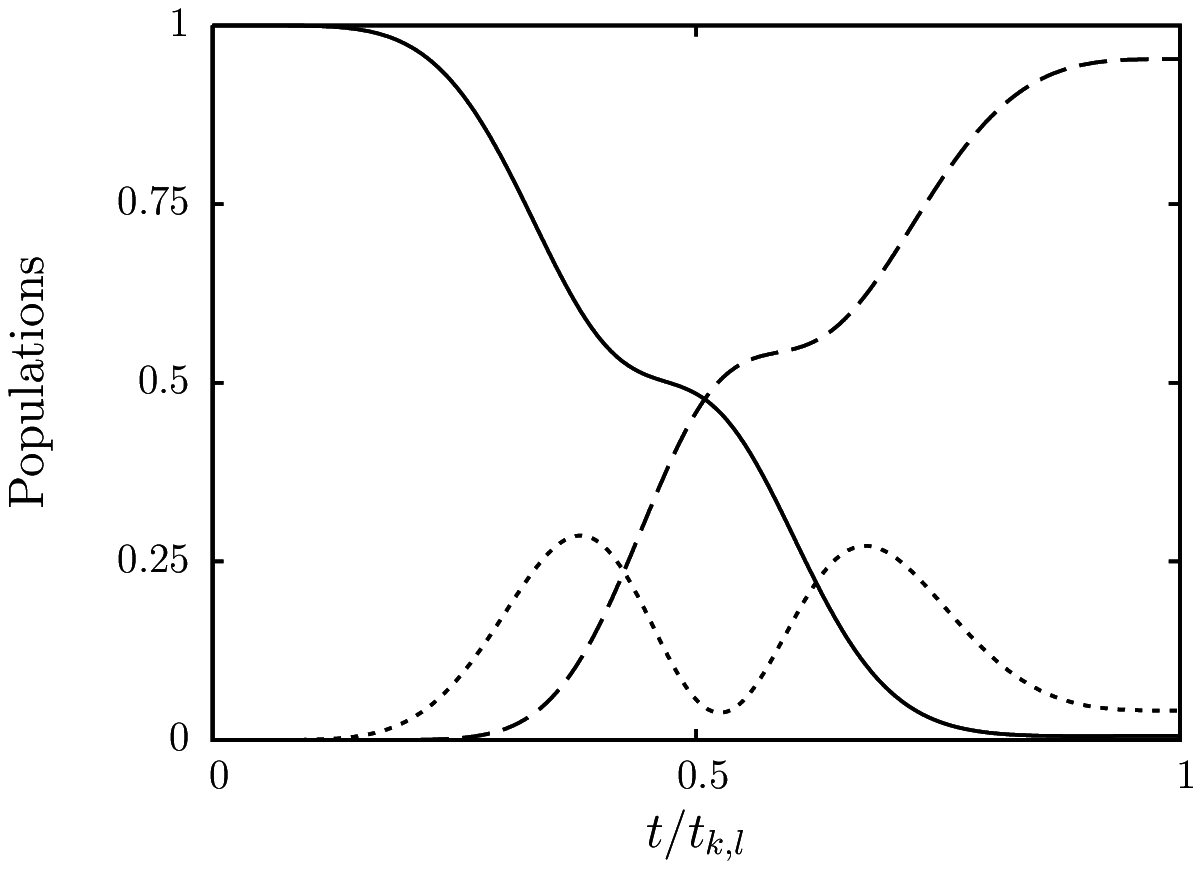}
  \caption{Populations of the resulting states for pulse
    excitation with the trapezium pulse (upper figure) and sine square
    pulse (lower figure) for $k=3$, $l=2$, $\kappa=0$, and
    $\gamma=0$. The meaning of the lines is the same as in
    Fig.~\ref{fig:4}}
  \label{fig:imp1}
\end{figure}

To illustrate this situation we plot the evolution of the final states
populations for two different pulse shapes. In Fig.~\ref{fig:imp1} we
present the evolution of final state populations for the trapezium
shape with $t_{r}/t_{p}=t_{f}/t_{p}=0.1$ and the sine square pulse of
the same duration.  The other parameters are chosen as $k=3$, $l=2$,
$\kappa=0$, $\gamma=0$, and the pulse duration $t_{p}=t_{k,l}$ with
$t_{k,l}$ calculated with $\Omega'$ adjusted according
to~(\ref{eq:36}).
The small value of $k=3$ gives the detuning $\Delta_{3,2}/(2g)=1.2649$
which is rather small, and in this case, as is seen from
Fig.~\ref{fig:imp1}, the desired state is not of good quality. The
fidelity, assuming the pulse duration $t_{p}=t_{k,l}$, is equal to
$0.8176$ for the trapezium pulse and to $0.9529$ for the sine square
pulse. This means that the approximation for $t_{k,l}$ is not very
good for small detunings. Small detunings mean the situation is close to
resonance, and the solutions should be close to periodic
solutions~(\ref{eq:31}). Actually, the oscillations are still visible in
the figure. The fidelities of the generated states can be improved by
adjusting the pulse duration. If the pulse duration is increased by a
factor of $1.103$, for the trapezium pulse, we get the fidelity equal
to $0.9681$, and this is the best result for the trapezium with
$\Delta_{3,2}$. In the case of the sine square pulse, by increasing the
pulse length by a factor of $1.087$, we get a fidelity better than
$0.99999$. In this respect, for small detunings, the sine square pulse
appears to be better than the trapezium pulse. We have to remember,
however, that the results refer to the ideal situation, where there is
no damping.

\begin{figure}[h]
  \centering
  \includegraphics[width=0.4\textwidth]{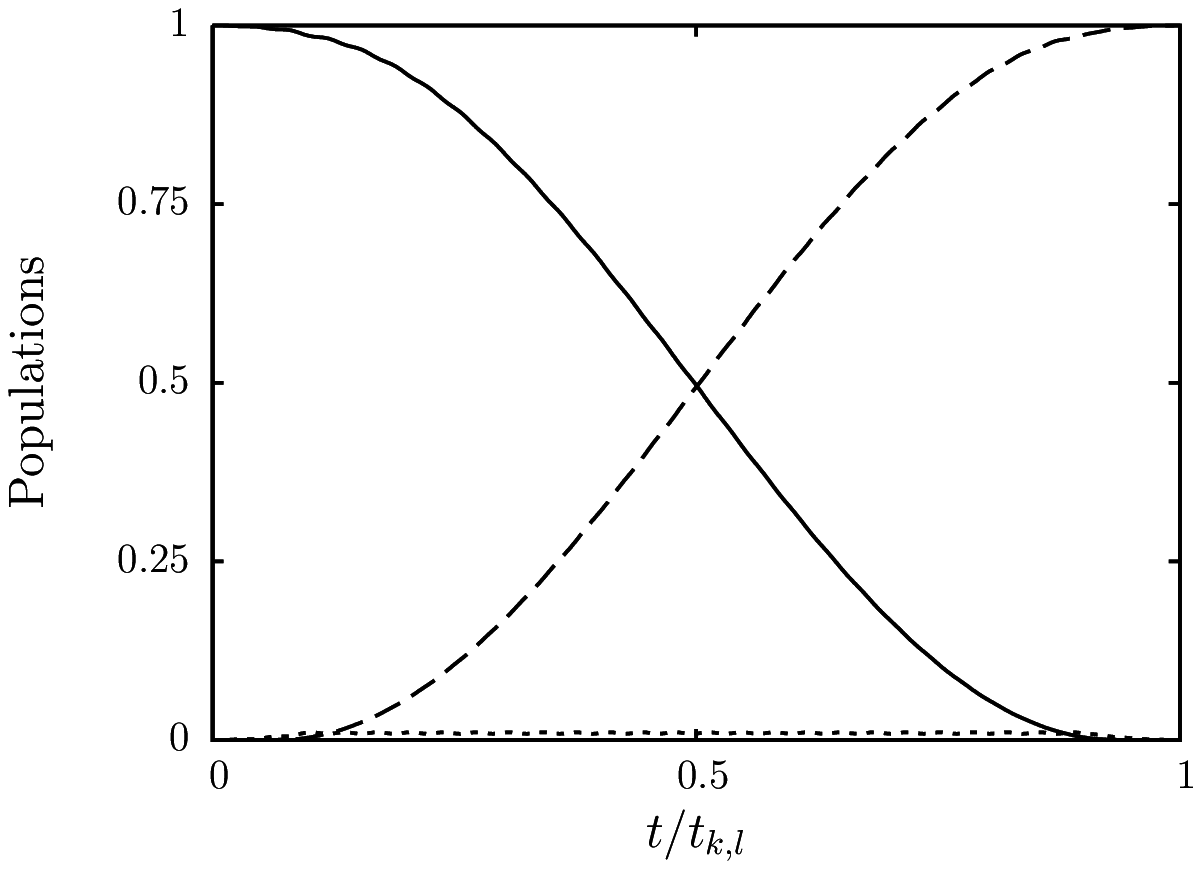}
  \includegraphics[width=0.4\textwidth]{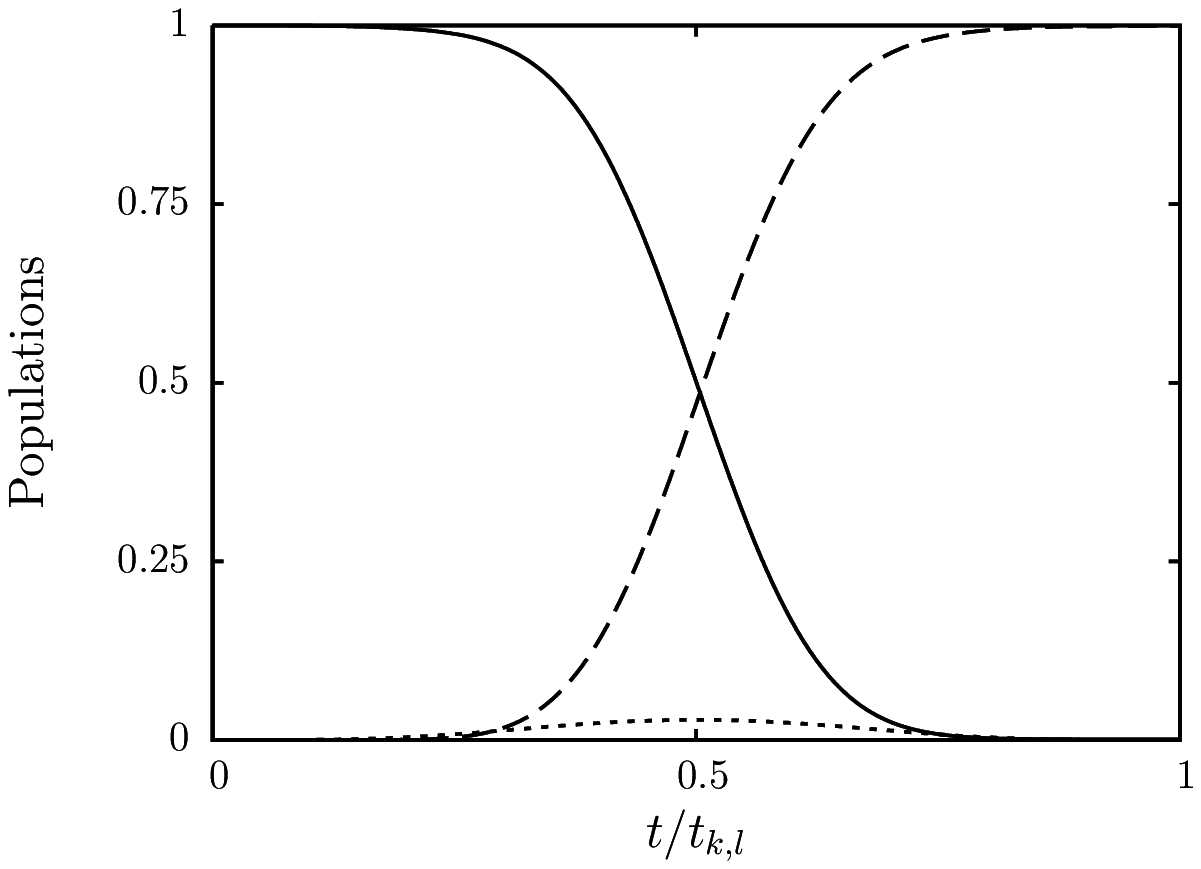}
  \caption{Same as Fig.~\ref{fig:imp1}, but for $k=31$ and $l=2$.}
  \label{fig:imp2}
\end{figure}

In Fig.~\ref{fig:imp2} we illustrate the situation for $k=31$ and
$l=2$.
In this case we have detuning $\Delta_{31,2}/(2g)=5.4321$, and the pulse
duration time $t_{p}=t_{k,l}$ calculated according to the above procedure
fits much better for producing desired state. The fidelities obtained
with such $t_{p}$ are: $0.99994$ for the trapezium shape and $0.99899$
for the sine square pulse. The results are remarkable and confirm that
the procedure works pretty well for larger detunings, i.e., higher
values of $k$. The result for the trapezium pulse is optimal for the
detuning $\Delta_{31,2}$, while the result for the sine square pulse can be
still improved when the pulse duration $t_{k,l}$ is multiplied by a
factor of $1.02$ giving a fidelity better than $0.99999$. Since the
evolution is not periodic as it is in the case of rectangular pulses, the
procedure will not work so well for higher values of $l$. The results
obtained give evidence that for pulse excitation the idea of fine
tuning of the quantum operations is also applicable.

In a real experimental situation, we have to take into account the
damping rates $\kappa$ and $\gamma$. In analogy to the results for
rectangular pulses, we can expect that the pulse duration should be
increased to get desired states with good quality. The approximation
described here does not give any hint as to how long the pulse should be,
but starting from the ideal case it is easy to adjust the time
numerically. For example, for the sine square pulse, taking
$\kappa/(2g)=0.05$, $\gamma/(2g)=0.03$, and other parameters as in
Fig.~\ref{fig:imp2}, we get a fidelity better than $0.99999$ for the
time $1.154\, t_{k,l}$. Even in this case the approximate value of the
pulse duration $t_{p}=t_{k,l}$ is a good starting point for numerical
optimization.

\section{Conclusion}
In conclusion, we have studied the effect of the population of the
intermediate state on two basic operations performed via the
two-photon Raman transition in the atomic $\Lambda$ system.  Full
dynamics of the three-level system is taken into account without
adiabatic elimination of the intermediate level. The dynamics studied
are conditioned on no photon detection emitted by spontaneous emission
and no photon detection leaking out of the cavity.  We have found a
discrete set of detunings leading to perfect $\pi$ pulse and $\pi/2$
pulse operations. It is shown how to make the population
oscillations for improving the quantum operation precision useful. In the
ideal case of $\kappa=0$ and $\gamma=0$, the population of the
intermediate state $|2\rangle$ oscillates, becoming periodically zero
and giving perfect operations.  For nonzero, but small values of
$\kappa$ and $\gamma$, the fidelity can achieve values very close to
unity. Approximate analytical solutions obtained in first order
perturbation theory are given, including both the cavity damping rate
$\kappa$ as well as the spontaneous emission rate $\gamma$. Numerical
results show that at certain circumstances, the spontaneous emission
can play a positive, stabilizing role and improve the performance of
quantum operations. The 
idea of fine tuning can also be applied for the pulse excitation with
pulse shapes different than rectangle.

For realistic, moderate values of the detuning, the intermediate state
has significant influence on the reliability of the quantum operation, but its
destructive role can be limited by the fine tuning presented here.
The possibility of using small or moderate values of detunings,
without reducing the operation fidelity by the nonzero population of
the excited state, is also important because the operation time is a
function of the detuning, and short operation times are preferred to
minimize effects of dissipation.


\begin{thebibliography}{31}
  \expandafter\ifx\csname
  natexlab\endcsname\relax\def\natexlab#1{#1}\fi
  \expandafter\ifx\csname bibnamefont\endcsname\relax
  \def\bibnamefont#1{#1}\fi \expandafter\ifx\csname
  bibfnamefont\endcsname\relax \def\bibfnamefont#1{#1}\fi
  \expandafter\ifx\csname citenamefont\endcsname\relax
  \def\citenamefont#1{#1}\fi \expandafter\ifx\csname
  url\endcsname\relax \def\url#1{\texttt{#1}}\fi
  \expandafter\ifx\csname urlprefix\endcsname\relax\def\urlprefix{URL
  }\fi \providecommand{\bibinfo}[2]{#2}
  \providecommand{\eprint}[2][]{\url{#2}}

\bibitem[{\citenamefont{Langer et~al.}(2005)\citenamefont{Langer,
      Ozeri, Jost, Chiaverini, DeMarco, Ben-Kish, Blakestad, Britton,
      Hume, Itano et~al.}}]{langer05_10s}
  \bibinfo{author}{\bibfnamefont{C.}~\bibnamefont{Langer}},
  \bibinfo{author}{\bibfnamefont{R.}~\bibnamefont{Ozeri}},
  \bibinfo{author}{\bibfnamefont{J.~D.} \bibnamefont{Jost}},
  \bibinfo{author}{\bibfnamefont{J.}~\bibnamefont{Chiaverini}},
  \bibinfo{author}{\bibfnamefont{B.}~\bibnamefont{DeMarco}},
  \bibinfo{author}{\bibfnamefont{A.}~\bibnamefont{Ben-Kish}},
  \bibinfo{author}{\bibfnamefont{R.~B.} \bibnamefont{Blakestad}},
  \bibinfo{author}{\bibfnamefont{J.}~\bibnamefont{Britton}},
  \bibinfo{author}{\bibfnamefont{D.~B.} \bibnamefont{Hume}},
  \bibinfo{author}{\bibfnamefont{W.~M.} \bibnamefont{Itano}},
  \bibnamefont{et~al.}, \bibinfo{journal}{Phys. Rev. Lett.}
  \textbf{\bibinfo{volume}{95}}, \bibinfo{pages}{060502}
  (\bibinfo{year}{2005}).

\bibitem[{\citenamefont{Pellizzari
      et~al.}(1995)\citenamefont{Pellizzari, Gardiner, Cirac, and
      Zoller}}]{pell95:_decoh}
  \bibinfo{author}{\bibfnamefont{T.}~\bibnamefont{Pellizzari}},
  \bibinfo{author}{\bibfnamefont{S.~A.} \bibnamefont{Gardiner}},
  \bibinfo{author}{\bibfnamefont{J.~I.} \bibnamefont{Cirac}},
  \bibnamefont{and}
  \bibinfo{author}{\bibfnamefont{P.}~\bibnamefont{Zoller}},
  \bibinfo{journal}{Phys. Rev. Lett.} \textbf{\bibinfo{volume}{75}},
  \bibinfo{pages}{3788} (\bibinfo{year}{1995}).

\bibitem[{\citenamefont{Bose et~al.}(1999)\citenamefont{Bose, Knight,
      Plenio, and Vedral}}]{bose}
  \bibinfo{author}{\bibfnamefont{S.}~\bibnamefont{Bose}},
  \bibinfo{author}{\bibfnamefont{P.~L.} \bibnamefont{Knight}},
  \bibinfo{author}{\bibfnamefont{M.~B.} \bibnamefont{Plenio}},
  \bibnamefont{and}
  \bibinfo{author}{\bibfnamefont{V.}~\bibnamefont{Vedral}},
  \bibinfo{journal}{Phys. Rev. Lett.} \textbf{\bibinfo{volume}{83}},
  \bibinfo{pages}{5158} (\bibinfo{year}{1999}).

\bibitem[{\citenamefont{Beige et~al.}(2000)\citenamefont{Beige, Braun,
      Tregenna, and Knight}}]{beige}
  \bibinfo{author}{\bibfnamefont{A.}~\bibnamefont{Beige}},
  \bibinfo{author}{\bibfnamefont{D.}~\bibnamefont{Braun}},
  \bibinfo{author}{\bibfnamefont{B.}~\bibnamefont{Tregenna}},
  \bibnamefont{and} \bibinfo{author}{\bibfnamefont{P.~L.}
    \bibnamefont{Knight}}, \bibinfo{journal}{Phys. Rev. Lett.}
  \textbf{\bibinfo{volume}{85}}, \bibinfo{pages}{1762}
  (\bibinfo{year}{2000}).

\bibitem[{\citenamefont{Chimczak}(2005)}]{chimczak:_entanglement}
  \bibinfo{author}{\bibfnamefont{G.}~\bibnamefont{Chimczak}},
  \bibinfo{journal}{Phys. Rev. A} \textbf{\bibinfo{volume}{71}},
  \bibinfo{pages}{052305} (\bibinfo{year}{2005}).

\bibitem[{\citenamefont{Lim et~al.}(2006)\citenamefont{Lim, Barrett,
      Beige, Kok, and Kwek}}]{lim06:_repeat}
  \bibinfo{author}{\bibfnamefont{Y.~L.} \bibnamefont{Lim}},
  \bibinfo{author}{\bibfnamefont{S.~D.} \bibnamefont{Barrett}},
  \bibinfo{author}{\bibfnamefont{A.}~\bibnamefont{Beige}},
  \bibinfo{author}{\bibfnamefont{P.}~\bibnamefont{Kok}},
  \bibnamefont{and} \bibinfo{author}{\bibfnamefont{L.~C.}
    \bibnamefont{Kwek}}, \bibinfo{journal}{Phys. Rev. A}
  \textbf{\bibinfo{volume}{73}}, \bibinfo{pages}{012304}
  (\bibinfo{year}{2006}).

\bibitem[{\citenamefont{Yuan and Zhu}(2007)}]{yuanJPB07}
  \bibinfo{author}{\bibfnamefont{C.-H.} \bibnamefont{Yuan}}
  \bibnamefont{and} \bibinfo{author}{\bibfnamefont{K.-D.}
    \bibnamefont{Zhu}}, \bibinfo{journal}{J. Phys. B}
  \textbf{\bibinfo{volume}{40}}, \bibinfo{pages}{801}
  (\bibinfo{year}{2007}).

\bibitem[{\citenamefont{Parkins et~al.}(1993)\citenamefont{Parkins,
      Marte, Zoller, and Kimble}}]{parkins93:_synth_zeeman}
  \bibinfo{author}{\bibfnamefont{A.~S.} \bibnamefont{Parkins}},
  \bibinfo{author}{\bibfnamefont{P.}~\bibnamefont{Marte}},
  \bibinfo{author}{\bibfnamefont{P.}~\bibnamefont{Zoller}},
  \bibnamefont{and} \bibinfo{author}{\bibfnamefont{H.~J.}
    \bibnamefont{Kimble}}, \bibinfo{journal}{Phys. Rev. Lett.}
  \textbf{\bibinfo{volume}{71}}, \bibinfo{pages}{3095}
  (\bibinfo{year}{1993}).

\bibitem[{\citenamefont{Riebe et~al.}(2004)\citenamefont{Riebe,
      H\"affner, Roos, H\"ansel, Benhelm, Lancaster, K\"orber, Becher,
      Schmidt-Kaler, James et~al.}}]{riebe04}
  \bibinfo{author}{\bibfnamefont{M.}~\bibnamefont{Riebe}},
  \bibinfo{author}{\bibfnamefont{H.}~\bibnamefont{H\"affner}},
  \bibinfo{author}{\bibfnamefont{C.~F.} \bibnamefont{Roos}},
  \bibinfo{author}{\bibfnamefont{W.}~\bibnamefont{H\"ansel}},
  \bibinfo{author}{\bibfnamefont{J.}~\bibnamefont{Benhelm}},
  \bibinfo{author}{\bibfnamefont{G.~P.~T.} \bibnamefont{Lancaster}},
  \bibinfo{author}{\bibfnamefont{T.~W.} \bibnamefont{K\"orber}},
  \bibinfo{author}{\bibfnamefont{C.}~\bibnamefont{Becher}},
  \bibinfo{author}{\bibfnamefont{F.}~\bibnamefont{Schmidt-Kaler}},
  \bibinfo{author}{\bibfnamefont{D.~F.~V.} \bibnamefont{James}},
  \bibnamefont{et~al.}, \bibinfo{journal}{Nature~(London)}
  \textbf{\bibinfo{volume}{429}}, \bibinfo{pages}{734}
  (\bibinfo{year}{2004}).

\bibitem[{\citenamefont{Barrett et~al.}(2004)\citenamefont{Barrett,
      Chiaverini, Schaetz, Britton, Itano, Jost, Knill, Langer,
      Leibfried, Ozeri et~al.}}]{barrett04}
  \bibinfo{author}{\bibfnamefont{M.~D.} \bibnamefont{Barrett}},
  \bibinfo{author}{\bibfnamefont{J.}~\bibnamefont{Chiaverini}},
  \bibinfo{author}{\bibfnamefont{T.}~\bibnamefont{Schaetz}},
  \bibinfo{author}{\bibfnamefont{J.}~\bibnamefont{Britton}},
  \bibinfo{author}{\bibfnamefont{W.~M.} \bibnamefont{Itano}},
  \bibinfo{author}{\bibfnamefont{J.~D.} \bibnamefont{Jost}},
  \bibinfo{author}{\bibfnamefont{E.}~\bibnamefont{Knill}},
  \bibinfo{author}{\bibfnamefont{C.}~\bibnamefont{Langer}},
  \bibinfo{author}{\bibfnamefont{D.}~\bibnamefont{Leibfried}},
  \bibinfo{author}{\bibfnamefont{R.}~\bibnamefont{Ozeri}},
  \bibnamefont{et~al.}, \bibinfo{journal}{Nature~(London)}
  \textbf{\bibinfo{volume}{429}}, \bibinfo{pages}{737}
  (\bibinfo{year}{2004}).

\bibitem[{\citenamefont{McKeever et~al.}(2004)\citenamefont{McKeever,
      Boca, Boozer, Miller, Buck, Kuzmich, and
      Kimble}}]{mckeever04:_single_photon}
  \bibinfo{author}{\bibfnamefont{J.}~\bibnamefont{McKeever}},
  \bibinfo{author}{\bibfnamefont{A.}~\bibnamefont{Boca}},
  \bibinfo{author}{\bibfnamefont{A.~D.} \bibnamefont{Boozer}},
  \bibinfo{author}{\bibfnamefont{R.}~\bibnamefont{Miller}},
  \bibinfo{author}{\bibfnamefont{J.~R.} \bibnamefont{Buck}},
  \bibinfo{author}{\bibfnamefont{A.}~\bibnamefont{Kuzmich}},
  \bibnamefont{and} \bibinfo{author}{\bibfnamefont{H.~J.}
    \bibnamefont{Kimble}}, \bibinfo{journal}{Science}
  \textbf{\bibinfo{volume}{303}}, \bibinfo{pages}{1992}
  (\bibinfo{year}{2004}).

\bibitem[{\citenamefont{Boozer et~al.}(2007)\citenamefont{Boozer,
      Boca, Miller, Northup, and Kimble}}]{boozerPRL07_map}
  \bibinfo{author}{\bibfnamefont{A.~D.} \bibnamefont{Boozer}},
  \bibinfo{author}{\bibfnamefont{A.}~\bibnamefont{Boca}},
  \bibinfo{author}{\bibfnamefont{R.}~\bibnamefont{Miller}},
  \bibinfo{author}{\bibfnamefont{T.~E.} \bibnamefont{Northup}},
  \bibnamefont{and} \bibinfo{author}{\bibfnamefont{H.~J.}
    \bibnamefont{Kimble}}, \bibinfo{journal}{Phys. Rev. Lett.}
  \textbf{\bibinfo{volume}{98}}, \bibinfo{pages}{193601}
  (\bibinfo{year}{2007}).

\bibitem[{\citenamefont{Legero et~al.}(2004)\citenamefont{Legero,
      Wilk, Hennrich, Rempe, and Kuhn}}]{legero04}
  \bibinfo{author}{\bibfnamefont{T.}~\bibnamefont{Legero}},
  \bibinfo{author}{\bibfnamefont{T.}~\bibnamefont{Wilk}},
  \bibinfo{author}{\bibfnamefont{M.}~\bibnamefont{Hennrich}},
  \bibinfo{author}{\bibfnamefont{G.}~\bibnamefont{Rempe}},
  \bibnamefont{and}
  \bibinfo{author}{\bibfnamefont{A.}~\bibnamefont{Kuhn}},
  \bibinfo{journal}{Phys. Rev. Lett.} \textbf{\bibinfo{volume}{93}},
  \bibinfo{pages}{070503} (\bibinfo{year}{2004}).

\bibitem[{\citenamefont{Hijlkema et~al.}(2007)\citenamefont{Hijlkema,
      Weber, Specht, Webster, Kuhn, and
      Rempe}}]{hijlkemaNatPhy07_1fotSer}
  \bibinfo{author}{\bibfnamefont{M.}~\bibnamefont{Hijlkema}},
  \bibinfo{author}{\bibfnamefont{B.}~\bibnamefont{Weber}},
  \bibinfo{author}{\bibfnamefont{H.~P.} \bibnamefont{Specht}},
  \bibinfo{author}{\bibfnamefont{S.~C.} \bibnamefont{Webster}},
  \bibinfo{author}{\bibfnamefont{A.}~\bibnamefont{Kuhn}},
  \bibnamefont{and}
  \bibinfo{author}{\bibfnamefont{G.}~\bibnamefont{Rempe}},
  \bibinfo{journal}{Nat. Phys.} \textbf{\bibinfo{volume}{3}},
  \bibinfo{pages}{253} (\bibinfo{year}{2007}).

\bibitem[{\citenamefont{Preskill}(1998)}]{preskill}
  \bibinfo{author}{\bibfnamefont{J.}~\bibnamefont{Preskill}},
  \bibinfo{journal}{Proc. R. Soc. London, Ser. A}
  \textbf{\bibinfo{volume}{454}}, \bibinfo{pages}{385}
  (\bibinfo{year}{1998}).

\bibitem[{\citenamefont{Steane}(1999)}]{steane99_doklad}
  \bibinfo{author}{\bibfnamefont{A.~M.} \bibnamefont{Steane}},
  \bibinfo{journal}{Nature~(London)} \textbf{\bibinfo{volume}{399}},
  \bibinfo{pages}{124} (\bibinfo{year}{1999}).

\bibitem[{\citenamefont{van Enk et~al.}(1997)\citenamefont{van Enk,
      Cirac, and Zoller}}]{Enk97}
  \bibinfo{author}{\bibfnamefont{S.~J.} \bibnamefont{van Enk}},
  \bibinfo{author}{\bibfnamefont{J.~I.} \bibnamefont{Cirac}},
  \bibnamefont{and}
  \bibinfo{author}{\bibfnamefont{P.}~\bibnamefont{Zoller}},
  \bibinfo{journal}{Phys. Rev. Lett.} \textbf{\bibinfo{volume}{78}},
  \bibinfo{pages}{4293} (\bibinfo{year}{1997}).

\bibitem[{\citenamefont{Pachos and Walther}(2002)}]{pachos02}
  \bibinfo{author}{\bibfnamefont{J.}~\bibnamefont{Pachos}}
  \bibnamefont{and}
  \bibinfo{author}{\bibfnamefont{H.}~\bibnamefont{Walther}},
  \bibinfo{journal}{Phys. Rev. Lett.} \textbf{\bibinfo{volume}{89}},
  \bibinfo{pages}{187903} (\bibinfo{year}{2002}).

\bibitem[{\citenamefont{Sch\"on et~al.}(2007)\citenamefont{Sch\"on,
      Hammerer, Wolf, Cirac, and Solano}}]{schonPRA07}
  \bibinfo{author}{\bibfnamefont{C.}~\bibnamefont{Sch\"on}},
  \bibinfo{author}{\bibfnamefont{K.}~\bibnamefont{Hammerer}},
  \bibinfo{author}{\bibfnamefont{M.~M.} \bibnamefont{Wolf}},
  \bibinfo{author}{\bibfnamefont{J.~I.} \bibnamefont{Cirac}},
  \bibnamefont{and}
  \bibinfo{author}{\bibfnamefont{E.}~\bibnamefont{Solano}},
  \bibinfo{journal}{Phys. Rev. A} \textbf{\bibinfo{volume}{75}},
  \bibinfo{pages}{032311} (\bibinfo{year}{2007}).

\bibitem[{\citenamefont{Kiraz et~al.}(2004)\citenamefont{Kiraz,
      Atat\"ure, and Imamo\=glu}}]{kirazPRA04}
  \bibinfo{author}{\bibfnamefont{A.}~\bibnamefont{Kiraz}},
  \bibinfo{author}{\bibfnamefont{M.}~\bibnamefont{Atat\"ure}},
  \bibnamefont{and}
  \bibinfo{author}{\bibfnamefont{A.}~\bibnamefont{Imamo\=glu}},
  \bibinfo{journal}{Phys. Rev. A} \textbf{\bibinfo{volume}{69}},
  \bibinfo{pages}{032305} (\bibinfo{year}{2004}).

\bibitem[{\citenamefont{Djuric and Search}(2007)}]{djuricPRB07}
  \bibinfo{author}{\bibfnamefont{I.}~\bibnamefont{Djuric}}
  \bibnamefont{and} \bibinfo{author}{\bibfnamefont{C.~P.}
    \bibnamefont{Search}}, \bibinfo{journal}{Phys. Rev. B}
  \textbf{\bibinfo{volume}{75}}, \bibinfo{pages}{155307}
  (\bibinfo{year}{2007}).

\bibitem[{\citenamefont{Zhou et~al.}(2002)\citenamefont{Zhou, Chu, and
      Han}}]{zhou02_SQUIDLam}
  \bibinfo{author}{\bibfnamefont{Z.}~\bibnamefont{Zhou}},
  \bibinfo{author}{\bibfnamefont{Shih-I.} \bibnamefont{Chu}},
  \bibnamefont{and}
  \bibinfo{author}{\bibfnamefont{S.}~\bibnamefont{Han}},
  \bibinfo{journal}{Phys. Rev. B} \textbf{\bibinfo{volume}{66}},
  \bibinfo{pages}{054527} (\bibinfo{year}{2002}).

\bibitem[{\citenamefont{Amin et~al.}(2003)\citenamefont{Amin, Smirnov,
      and van~den Brink}}]{amin03_SQUIDLam}
  \bibinfo{author}{\bibfnamefont{M.~H.~S.} \bibnamefont{Amin}},
  \bibinfo{author}{\bibfnamefont{A.~Y.} \bibnamefont{Smirnov}},
  \bibnamefont{and} \bibinfo{author}{\bibfnamefont{A.~Maassen}
    \bibnamefont{van~den Brink}}, \bibinfo{journal}{Phys. Rev. B}
  \textbf{\bibinfo{volume}{67}}, \bibinfo{pages}{100508(R)}
  (\bibinfo{year}{2003}).

\bibitem[{\citenamefont{Yang et~al.}(2003)\citenamefont{Yang, Chu, and
      Han}}]{yang03_SQUIDLam} \bibinfo{author}{\bibfnamefont{C.-P.}
    \bibnamefont{Yang}}, \bibinfo{author}{\bibfnamefont{Shih-I.}
    \bibnamefont{Chu}}, \bibnamefont{and}
  \bibinfo{author}{\bibfnamefont{S.}~\bibnamefont{Han}},
  \bibinfo{journal}{Phys. Rev. A} \textbf{\bibinfo{volume}{67}},
  \bibinfo{pages}{042311} (\bibinfo{year}{2003}).

\bibitem[{\citenamefont{Yang and Han}(2004)}]{yang04_SQUIDLam}
  \bibinfo{author}{\bibfnamefont{C.-P.} \bibnamefont{Yang}}
  \bibnamefont{and}
  \bibinfo{author}{\bibfnamefont{S.}~\bibnamefont{Han}},
  \bibinfo{journal}{Phys. Lett. A} \textbf{\bibinfo{volume}{321}},
  \bibinfo{pages}{273} (\bibinfo{year}{2004}).

\bibitem[{\citenamefont{Yang et~al.}(2004)\citenamefont{Yang, Chu, and
      Han}}]{yangPRL04_SQUIDLam} \bibinfo{author}{\bibfnamefont{C.-P.}
    \bibnamefont{Yang}}, \bibinfo{author}{\bibfnamefont{Shih-I.}
    \bibnamefont{Chu}}, \bibnamefont{and}
  \bibinfo{author}{\bibfnamefont{S.}~\bibnamefont{Han}},
  \bibinfo{journal}{Phys. Rev. Lett.} \textbf{\bibinfo{volume}{92}},
  \bibinfo{pages}{117902} (\bibinfo{year}{2004}).

\bibitem[{\citenamefont{Tregenna et~al.}(2002)\citenamefont{Tregenna,
      Beige, and Knight}}]{tregenna02_cnot}
  \bibinfo{author}{\bibfnamefont{B.}~\bibnamefont{Tregenna}},
  \bibinfo{author}{\bibfnamefont{A.}~\bibnamefont{Beige}},
  \bibnamefont{and} \bibinfo{author}{\bibfnamefont{P.~L.}
    \bibnamefont{Knight}}, \bibinfo{journal}{Phys. Rev. A}
  \textbf{\bibinfo{volume}{65}}, \bibinfo{pages}{032305}
  (\bibinfo{year}{2002}).

\bibitem[{\citenamefont{Goto and Ichimura}(2004)}]{gotoPRA04}
  \bibinfo{author}{\bibfnamefont{H.}~\bibnamefont{Goto}}
  \bibnamefont{and}
  \bibinfo{author}{\bibfnamefont{K.}~\bibnamefont{Ichimura}},
  \bibinfo{journal}{Phys. Rev. A} \textbf{\bibinfo{volume}{70}},
  \bibinfo{pages}{012305} (\bibinfo{year}{2004}).

\bibitem[{\citenamefont{Kis and
      Renzoni}(2002)}]{kis02:_qubit_rotat_by_stimul_raman_adiab_passag}
  \bibinfo{author}{\bibfnamefont{Z.}~\bibnamefont{Kis}}
  \bibnamefont{and}
  \bibinfo{author}{\bibfnamefont{F.}~\bibnamefont{Renzoni}},
  \bibinfo{journal}{Phys. Rev. A} \textbf{\bibinfo{volume}{65}},
  \bibinfo{pages}{032318} (\bibinfo{year}{2002}).

\bibitem[{\citenamefont{Sangouard
      et~al.}(2005)\citenamefont{Sangouard, Lacour, Gu\'erin, and
      Jauslin}}]{sangouard05:_fast_swap_gate_by_adiab_passag}
  \bibinfo{author}{\bibfnamefont{N.}~\bibnamefont{Sangouard}},
  \bibinfo{author}{\bibfnamefont{X.}~\bibnamefont{Lacour}},
  \bibinfo{author}{\bibfnamefont{S.}~\bibnamefont{Gu\'erin}},
  \bibnamefont{and} \bibinfo{author}{\bibfnamefont{H.~R.}
    \bibnamefont{Jauslin}}, \bibinfo{journal}{Phys. Rev. A}
  \textbf{\bibinfo{volume}{72}}, \bibinfo{pages}{062309}
  (\bibinfo{year}{2005}).

\bibitem[{\citenamefont{Chimczak et~al.}(2005)\citenamefont{Chimczak,
      Tana\'s, and Miranowicz}}]{chimczak:_entanglement_teleportation}
  \bibinfo{author}{\bibfnamefont{G.}~\bibnamefont{Chimczak}},
  \bibinfo{author}{\bibfnamefont{R.}~\bibnamefont{Tana\'s}},
  \bibnamefont{and}
  \bibinfo{author}{\bibfnamefont{A.}~\bibnamefont{Miranowicz}},
  \bibinfo{journal}{Phys. Rev. A} \textbf{\bibinfo{volume}{71}},
  \bibinfo{pages}{032316} (\bibinfo{year}{2005}).

\end{thebibliography}

\end{document}